\documentclass[doublespacing]{elsart}
\usepackage{graphicx}
\begin{document}
\begin{frontmatter}

\title{Andreev reflection and spin polarization measurement of Co/YBCO junction}

         % Enter your title between curly braces

\author{ N. Ghosh$^{1}$\corauthref{cor1}},\ead{ghosh.nilotpal@gmail.com }\corauth[cor1]{corresponding author}\author{ J. Barzola Quiquia$^{1}$, Q. Xu$^{2}$,  G. Biehne$^{1}$, H. Hochmuth$^{1}$,}
 \author{  M. Lorenz$^{1}$, P.Esquinazi$^{1}$, M. Grundmann$^{1}$ and  H. Schmidt$^{2}$ }

\address{$^{1}$Institut f{\"u}r Experimentelle Physik II, Fakult{\"a}t
f{\"u}r Physik und Geowissenschaften,\\
 Universit{\"a}t Leipzig, Germany,\\
 $^{2}$Forschungszentrum Dresden-Rossendorf, Institut f{\"u}r lonenstrahlphysik und Materialforschung, Germany}

\date{\today}          % Enter your date or \today between curly braces

%\newpage
\begin{abstract}

%\noindent{ \bf Abstract}
%\end{center}
\normalsize \baselineskip=18pt
%\vskip.02cm
%\pagenumbering{roman}

We report temperature dependent Andreev reflection  measurements of Co/ Y$_{1}$Ba$_{2}$Cu$_{3}$O$_{7-\delta}$ (YBCO) heterostructure samples with  junction areas of 1 $\mu$m diameter. Modelling of the 5-70~K conductivity data according to a modified Blonder-Tinkham-Klapwijk theory yields a spin polarization in Co film amounting to  34$\%$  which is  almost constant up to 70~K. The YBCO films  have been grown by pulsed laser deposition on sapphire substrates. The Co films are deposited  by thermal evaporation on YBCO. The film is characterized by powder X-ray diffraction measurements which shows YBCO is grown in (001) direction.The critical current density, 5 x 10$^{6}$ A/cm$^{2}$, in YBCO remains nearly constant after deposition of  Co at zero field and 77~K.
\end{abstract}
\end{frontmatter}
%\newpage
\section{Introduction}

Andreev reflection experiments across
normal conductor (N)  and superconductor (SC) junctions represent
an useful technique to explore various interesting physical properties like spin polarization of ferromagnets, gap anisotropy of superconductors, etc.
Several mechanisms  are involved in current transport into superconductor, when a bias voltage V  is applied across a clean N/SC point contact.
The electrons can pass from N into SC as
quasielectrons or holes at voltages higher than the superconducting gap $\it{\Delta}$, which relax into the Cooper-pair condensate over the charge relaxation distance. However, for
voltages lower than  $\it{\Delta}$  no quasiparticle states are available in the superconductor. Instead, the
current is converted directly into a supercurrent of Cooper
pairs, consisting of two electrons of charge $\it{e}$ with opposite
spin. This is accomplished by the reflection of a hole back
into the metal, a process first described by Andreev\cite{andreev}.
When a normal metal
is replaced by a  ferromagnetic metal (FM), a suppression of  Andreev reflection (AR) occurs.  For a FM/SC contact  the  process involves a coherent
interspin-subband transfer which is sensitive to
the relative electronic spin density of states  at Fermi energy ($E_{F}$). If the spin polarization $P$ is zero,  the AR is  not hindered. However, if $P$ amounts to 100$\%$ near $E_{F}$,  there are no spin-down states available
in the metal for the reflected hole and AR is  completely  suppressed\cite{soulen,upa}. Hence, determination of $P$ for a FM can be
carried out by measuring  the suppression of AR.\\
The spin polarization $P$ of a 3-d  transition metal FM depends on its electronic structure containing narrow $d$-band and broad $s$-bands at  $E_{F}$.  Co is an  interesting 3-d metal with the Curie temperature amounting to 1388~K\cite{wohlf}. Spin-resolved photoelectron spectroscopy measurements showed that Co has short range ferromagnetic order above  the Curie temperature\cite{schn}. The first experiment
of spin-polarized transport and tunneling in Co was performed by Tedro and
Meservey  where $P$  was reported  to be  34$\%$  at 0.4~K \cite{tedro}.
 Measurement of $P$ by AR spectroscopy  is limited by the $T_{c}$ of the used SC.
Soulen $\it{et\ al.}$  reported $P$ = 42$\%$ for Co by AR measurements using  a  Nb tip\cite{soulen}. Since  the superconducting transition temperature ($T_{c}$) for Nb is around 9~K,  $P$ was determined only below that temperature.
 The resolution
of AR spectroscopy strongly depends on the temperature
$\it{T}$ due to the broadening of carrier distribution  by k$_{B}$T  around the Fermi level. Observation of distinct spectral features requires the condition k$_{B}$T $\ll$
k$_{B}$$\theta_{D}$ ($\theta_{D}$ = Debye Temperature)  i.e. measurements should be taken at temperatures
considerably lower than the Debye temperature\cite{Nayduk}.
%Typically, AR is probed in the temperature range from 1.5 to 4.2~K.
%
In order to determine $P$ at higher temperature, the necessity of using high-T$_{c}$  SC as a  superconducting electrode for  AR measurements is necessary.
A dip in differential conductance of a ferromagnet/high-T$_{c}$ superconductor junction  has been  reported as a consequence of suppressed AR\cite{vasko}.
This indicates that the
detection of $\it{P}$ by AR measurements across FM/high-T$_{c}$
superconductor junction  should be possible.\\
Here, we report on AR measurements on  Co/YBCO with
junction area of  1 $\mu$m diameter  at 5-100~K. Our analysis and modelling of the sub-gap conductivity variation yields $\it{P}$  around 34$\%$ for Co in the temperature range from 5 to 70~K. To the best of  our knowledge, this is the first time that the  spin polarization of Co has been probed by  Andreev reflection spectroscopy at such elevated temperatures, up to 70~K.\\

\section{ Thin film growth and characterization}
The investigated YBCO  films have been grown
on r-plane oriented sapphire (Al$_{2}$O$_{3}$) substrates with
underlying CeO$_{2}$ buffer layer by pulsed laser deposition (PLD)
technique\cite{lorenz}.In general, a large oxygen partial pressure
during the PLD growth of YBCO on sapphire amounting to 0.16 mbar is
required to maintain the stoichiometry of YBCO  and to retain its
superconductivity.  The deposited thin YBCO films  are (001) axis
oriented, orthorhombic in structure and their thickness amounts to
230 $\pm$ 30 nm in each sample. The Co/YBCO heterostructure has been
prepared by thermal evaporation of Co on YBCO under a vacuum level of 3 x 10$^{-3}$ mbar.
Thickness of  the Co film  on YBCO is around 11 nm.  \\
 The crystalline quality of the samples
has been investigated by powder X-ray diffraction (XRD) measurements  before and after deposition of Co using the Cu-K$_{\alpha}$ line ($\lambda$= 0.15406 nm).
The  XRD spectra  of the investigated  Co/YBCO heterostructures  in Fig.~1
 confirm the (001)-axis orientation of YBCO on r-sapphire. \\

\section{ Transport experiment}

\subsection{Critical current density of YBCO}
The critical current density ($j_{C}$) determines the quality of a superconductor. It is essential to probe the
superconductivity of YBCO before and after deposition of  Co.  We used an inductive method developed to
determine $j_{C}$ of YBCO films on single-sided or double-sided
sapphire substrates\cite{hochmuth}. The magnetic shielding of the
YBCO film is measured in dependence of temperature and AC current in
the measuring coil. The measurement is carried out by placing  the
sample into liquid nitrogen to work below YBCO  superconduction transition temperature($T_{c}$)
amounting to  92 K. We have found that  the critical current density
for YBCO  amounts to $j_{C}$ = 3 x 10$^{6}$ A/cm$^{2}$. The $j_{C}$
measurements on  Co/YBCO heterostructures
reveal that the critical current density of YBCO remains unchanged. This result proves that the superconductivity of YBCO is preserved  after deposition of Co.\\
\subsection{Sample preparation and AR measurements}
 A SiN insulating layer of 50 nm thickness has been   deposited on the YBCO film by sputtering deposition method. A single hole of  1 $\mu$m diameter
has been created  in SiN  by focused ion beam (FIB) etching.
 The diameter  was  1 $\mu$m at the top of the hole and  at the bottom the diameter was much less. Basically, the hole has a conical  shape \cite{buhrman}. Hence, the diameter at Co/YBCO contact must be much less than  1 $\mu$m.
Finally,  Co films have been deposited on SiN by  thermal evaporation. Hence,  the Co film on SiN will be about 61 nm thick in the region of the hole.
It is known that this kind of hole creation  by FIB usually get affected by  Gallium (Ga) ion contamination.
We have also faced the similar problems and  for affected contacts the contact resistance used to be very high.
However, we have taken the sample with best contact which has lowest resistance and carried out the AR measurement.
The normal contact resistance  has been around 16 Kohm. We have  calculated the  approximate contact area  using the formula  G$_{n}$ $\simeq$ e$^{2}$/$\hbar$ k$_{F}^{2}$ $A'$  where k$_{F}$ is the Fermi wave vector and  the   effective contact area ($A'$)  amounts to  0.65 nm$^{2}$. %
The measuring electrodes are fabricated by  silver paste on the contact pads.
Two  contacts are made on Co pads, while the other two
contacts are made on YBCO pads (Fig.~2).
%
%The junction resistance lies in the k$\Omega$ range at room temperature.
%
We have carried out  current($I$) vs voltage($V$)
   measurements  in four-probe geometry on  Co/YBCO samples with  junction area of 1 $\mu$m diameter from 5 to 100~K.
The conductivity  has been calculated by numerical differentiation of  the  $I$ vs. $V$ data.
   The temperature dependent normalized conductivity  data
    are shown in Fig.~3. We have observed clear Andreev reflection and modelled the data up to 70~K for Co/YBCO  samples.
Above 70~K it was not possible to model the data because of large background noise.

\section{ Modelling}

We base our analysis and modelling the conductivity data on
the pioneering work by Blonder-Tinkham-Klapwijk (BTK) for AR in the ballistic
regime \cite{BTK}.
The BTK formula has already been used for a much larger contact area of 1x1
mm$^{2}$ in Ga$_{0.95}$Mn$_{0.05}$As/Ga junctions\cite{Bradeen}.
 Our present Co/YBCO samples  have  much smaller  contact area in the 1 $\mu$m$^{2}$ range.
The BTK formula is valid for
normal metal/$\it{s}$-wave superconductor junction. However, in the
present investigation, the high-T$_{c}$ $\it{d}$-wave superconductor
YBCO has been used\cite{wei}.
In contrast to $\it{s}$-wave
superconductor, here the tunneling spectra strongly depend on
 the tunneling direction with respect to the crystallographic  axis. A
tunneling theory for normal metal-insulator-$\it{d}$-wave
superconductor has been reported by Tanaka and Kashiwaya
\cite{tana_kashi}.
It has been shown by Barash $\it{et\
al.}$ \cite{Barash} that the character of the change in the order
parameter at the boundary of a $\it{d}$-wave superconductor with a
normal metal or insulator does not differ much from a junction with
a $\it{s}$-wave superconductor when the orientation of the normal to
the $\it{d}$-wave superconductor is along the principal
crystallographic axes.
Because the investigated Co films
are deposited  along the (001) direction, the current direction across the
junction is predominantly in (001) direction.
 Hence, we  used the BTK
formulas (Eq.~1 and 2) for  energy ($E$) $<$ $\Delta$ \cite{BTK}.
%
%Also Komissinski $\it{et\ al.}$ used
%BTK formula for superconductor/normal metal/high-T$_{c}$
%superconductor heterostructures. \cite{khom}

\begin{equation}
I_{NS}= 2e N v_{f}  \mathcal{A}  \int[f_{0}(E-eV)-f_{0}(E)][1+A(E)-B(E)]dE
\end{equation}
\begin{equation}
%\begin{center}
 A(E) = \frac{\Delta^{2}}{E^{2}+ ( \Delta^{2}-E^{2})(1 + 2Z^{2})},
 B(E) = 1- A(E)
%\end{center}
\end{equation}

Here, $I_{NS}$ denotes the current across the junction, $e$ the electronic charge, $N$
the density of states, $v_{f}$ the  Fermi velocity, $\mathcal{A}$ the junction
cross sectional area, f$_{0}$ the Fermi distribution function, $V$ the
applied voltage, $A(E)$ the probability of Andreev reflection and $B(E)$ the
probability of normal reflection.
We have followed Strijkers $\it{et\ al.}$\cite{Strijker} for expressions of $A(E)$ and $B(E)$ in out of sub-gap region ( $E > \Delta $)  which are given below.
\begin{equation}
A(E)= \frac {u_{01}^{2}v_{02}^{2}}{\gamma_{2}^{2}},
B(E) = \frac {(u_{02}^{2} - v_{02}^{2}) Z^{2}(1+Z^{2})}{\gamma_{2}^{2}}
\end{equation}

 Since Co is a 3-d transition metal ferromagnet,  the carrier spin
polarization should also be an important parameter to be
carefully considered. Incorporating both $\it{Z}$ and $\it{P}$ we
modified the BTK formula from Strijkers $\it{et\ al.}$ (see
Eq.~4)

\begin{equation}
I_{NS} = (1-P)I_{U} + P I_{P},
\end{equation}

with $I_{U}$ being the unpolarized current and $I_{P}$ the polarized current
across the junction. Here, we have also  used separate expression for normal reflection probability ($B_{p}$)
of polarized current at $E > \Delta $ as shown below.
\begin{equation}
B_{p}(E) =  \frac {(u_{02}^{2} - v_{02}^{2}) Z^{2}(1+Z^{2})}{\gamma_{3}^{2}}
\end{equation}

The probability of Andreev reflection ($A_{p}$) for polarized current is zero. The expressions of $u_{01}$, $v_{01}$, $u_{02}$, $v_{02}$,$\gamma_{2}$ and $\gamma_{3}$  can be found in the   mentioned reference \cite{Strijker}.\\
In order to take into account the interface
scattering and finite life time effects of quasi particles, we also
considered the additional broadening parameter $\it\Gamma$($E^{'} = E + i\Gamma$).
This phenomenological idea has been employed by several authors
before\cite{Srikanth,Szabo}.
We have used  this modified BTK formula including a superconducting
gap ($\it\Delta$),  barrier strength ($\it{Z}$),  spin
polarization ($\it{P}$) and  broadening parameter($\it\Gamma$) to
model the observed  data of sub-gap conductivity variation. The
modelled data are shown in Fig.~3 and  they agree with the
experimental data in the sub-gap region. The value of fitting
parameters are displayed in the  legends.
The influence of proximity effect has been neglected in the present model. Because, the  existence of  superconducting proximity layer in a material depends on the formation of cooper pairs and  ferromagnetic material (like Co) breaks the cooper pairs \cite{Strijker}. If cooper pairs can not be formed, proximity layer will not exist.
%
%There are two gap
%parameters incorporated into the fitting procedure, one for Andreev reflection and proximity
%effect $\it\Delta_{1}$,  the other for
%quasiparticle transport $\Delta_{2}$\cite{Strijker}.
%
%It is already known, that Cooper pairs from a SC can diffuse into a N in close proximity  and develop weakly %superconducting layer at the N/SC interface. Actually, this  proximity layer  always has lower $T_{c}$ and  lower  %$\it\Delta$ than those of the bulk.
%The AR process  at N/proximity-layer  interface occurs at bias voltage smaller than  superconducting gap ($\it\Delta_{1}$) of the proximity layer.
%However, quasiparticles  can enter  the SC  when the voltage is higher than the bulk gap ($\Delta_{2}$) of SC.

%However,the fitting has been started with same initial values for $\it\Delta_{1}$ and $\it\Delta_{2}$.
%The values of $\it{Z}$($\leq$0.4) , $\it{P}$(76$\%$) and $\it\Gamma$ ($\approx$
%4.6 meV ) of Zn(Co,Al)O/YBCO heterostructure at 15~K are reasonable.

\section{ AR results and analysis}
Fig.~4  shows the plots of AR measurements and fits for
Co/YBCO at all measured temperatures.
We have noticed asymmetry in the conductance plots and it is more at higher temperature.
 Although the d-wave parameter has been neglected because of contact geometry (001),  the d-wave features  are   still present. The asymmetry in conductance and peaks seen above T$_{c}$ region are  basically d-wave features.
 The present contact dimension amounts to 1 $\mu$m$^{2}$  at the top of the hole.  Hence, the current direction may have  some components  in x-y direction also. Since, delta is not same in x ,y and z direction for YBCO, that anisotropy will be experienced by the current.  Actually, it was reported  that $\Delta$ along (110) = 27 meV, $\Delta$ along (100) = 28-29 meV and $\Delta$ along (001) = 19 meV for YBCO \cite{wei}.This can be a possible explanation for observed asymmetry in the conductance.
 The fits at 50 K, 60 K, and 70 K are not following exactly the experimental data because the effect of non-ohmicity at higher energy and  higher temperature is not included into our model.
  This non-ohmicity  has occured  from the temperature effect and possibly due to interface inhomogeneity.
 In addition,  we can not completely ignore the effect of crossing the critical current locally.
 Because, there are dips in conductance  at 50, 60, 70K, which  are very pronounced and the fit can not follow  them. When the current at any area of the N/SC contact region  is locally higher than the critical current, that part of the contact transforms into normal state. As a result, an unusual peak in resistance or a dip in conductance measurement is observed. For the present case, the dips  are seen at around 25 meV, where $\Delta$  at those temperatures are  in between 19 to 14 meV.\\

 Modelled  AR data of Co/YBCO  junctions  with  contact area of 1 $\mu$m  diameter according to modified BTK theory, agree well with the experimental data (Fig.~4). The values of fitting parameter are
displayed in Fig.~4 also. The magnitude of $\it{P}$ at 10-70~K lies around
34$\%$  for Co and similar to that  determined at 0.4~K by
tunneling measurements\cite{tedro}.
The dependence of fitting parameters on temperature is displayed in Fig.~5.
 It is noticed that the magnitude of $\it{P}$ does not depend on temperature up to 70~K. This is expected because the Curie
 temperature of Co is 1388~K\cite{wohlf}.
 This trend is similar to that obtained for Fe by Mukhopadhyay $\it{et\ al.}$\cite{sourin}.
 We found  that $\it{Z}$  increases with temperature. We point out that  $\it{Z}$ does not only  describe  the interface potential for the Co/YBCO interface. In reality, $\it{Z}$ is defined by $\it{Z}_{eff}$ where $\it{Z}_{eff}$ = $\it{Z}_{i}$ +$\frac {(r-1)^{2}}{4r}$. Here, $\it{r}$ is the ratio of Fermi velocities of normal metal and superconductor and $\it{Z}_{i}$  defines the imperfectness of the interface\cite{BT}. The temperature dependence of $\it{Z}_{i}$ is not known exactly. For a  ferromagnetic metal and superconductor interface, $\it{r}$ depends on  the different velocities in the up and down spin band of the ferromagnet. The present temperature range of  AR measurement is much below the ferromagnetic transition temperature of Co. Thus, the ferromagnetic spin fluctuation and the decrease in exchange splitting with temperature may not be of much significance for Co. Hence, the change of $\it{Z}$ up to  70~K is presumably dependent on  only $\it{Z}_{i}$.
 %
 %However, we attribute the  slight increase of $\it{Z}$  as the sample temperature approaches   $T_{c}$  to purely thermal effects where two competitive influences of $\it{Z}_{i}$ and  $r$  are possibly involved.
 %
We note  that  $\it{{\Gamma}}$  increases with temperature and peaks when the the sample temperature approaches   $T_{c}$. The life time of  quasiparticles ($\it{{\tau}}$) is incorporated into this broadening parameter $\it{{\Gamma}}$ = $\frac{\hbar}{\tau}$\cite{plecenik}. It implies that the quasiparticle scattering rate increases with temperature and attains maximum value close to $T_{c}$.
 %
 %It is obvious that   the life time of quasiparticle  decreases as the sample temperature approaches   $T_{c}$.
 %
$\it\Delta$  decreases with temperature and is almost nearly equal to zero near $T_{c}$. Although YBCO is a high-T$_{c}$ superconductor, the temperature dependence of $\it\Delta$ follows BCS theory.  A similar   temperature dependence of $\it\Delta$ in Bi$_{2}$Sr$_{2}$CaCu$_{2}$O$_{8+ \delta}$ has  been reported by S. I. Vedeneev $\it{et\ al.}$\cite{vedeneev}.
In Fig.~6  the spin polarization $\it{P}$ has been plotted against $\it{Z}$. The experimental points have been fitted with a straight line and extrapolated up to $\it{Z}$ = 0. As a result, we have found the  value of  $\it{P}$ as around 33$\%$ at $\it{Z}$ = 0.\\
We have successfully modelled  AR data  for Co/YBCO samples with small interface area.
 Theoretical calculations reported the exchange
splitting for Co
 around 1.39 eV, density of states  at E$_{f}$$\uparrow$
 N(E$_{f}$ $\uparrow$) = 4.29 per electrons/atom Ry spin
 and density of states  at E$_{f}$$\downarrow$ N(E$_{f}$
$\downarrow$)= 11.32 per electrons/atom Ry spin\cite{batallan}.
 This
indicates a theoretical value of spin polarization
  around 45$\%$.  The experimentally found
$\it{P}$ for Co at 0.4~K is 34$\%$ as reported by Tedro and
Meservey\cite{tedro}. Here, we  corroborate  such value of 34$\%$ up to 70~K from AR measurements on a Co/YBCO sample with junction area  of 1 $\mu$m diameter.  \\
%In conclusion, we have  measured
 %sub-gap conductivity variations of  Co/YBCO at 5-100~K. We have observed clear signature of %Andreev reflection below 90~K. Conductivity changes are
%modelled at 5-70~K according to modified BTK formula. We have found 34$\%$ spin polarization  in Co for Co/YBCO sample with 1x1 $\mu$m$^{2}$  interface area up to 70~K.
\subsection*{Acknowledgement}
 Alexander von Humboldt  Stifftung (N.G) and Bundesministerium f{\"u}r Bildung und Forschung (Q.X, G.B, H.H, H.S) are gratefully acknowledged for financial support. Preparation of the junction  contact on the Co/YBCO sample  has been carried out using a dual beam microscope FEI Nanolab 200XT supported by HBFG036-371. This work was partially supported by DFG  with in the framework of SFB 762. We thank  Dr. V. Gottschalch for providing the facility for SiN deposition and G. Ramm for PLD target preparation.

\newpage
\subsection*{ Figure and Figure Captions}

\begin{figure*}[h]
\includegraphics[width=10cm]{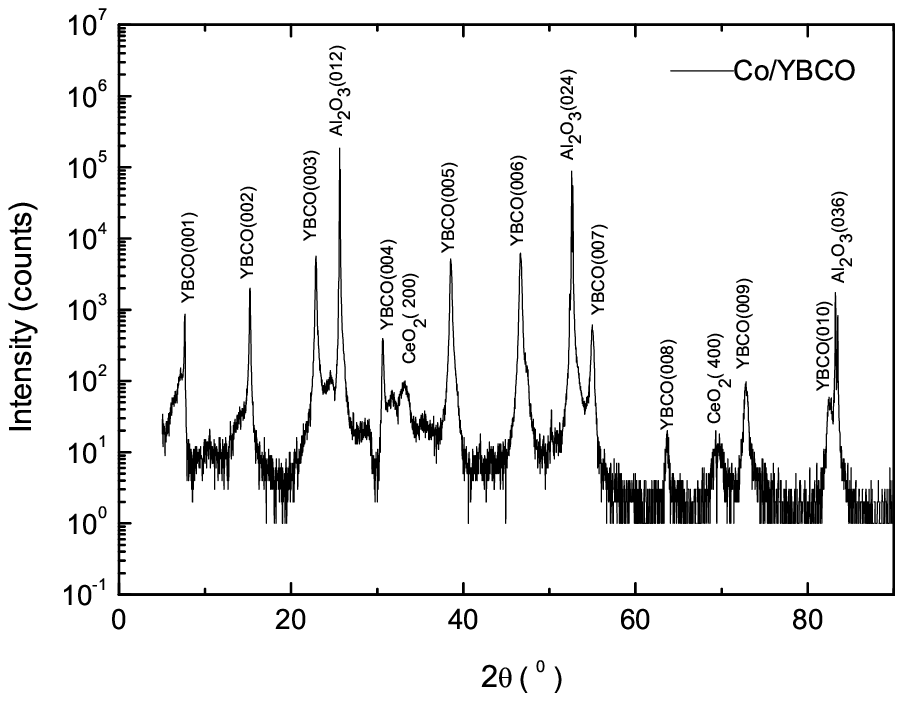}
\caption{\label{fig:epsart}2$\theta-\omega$ X-ray diffraction
patterns measured on a Co/YBCO  thin film  on CeO$_{2}$-buffered r-plane sapphire
substrate. No Co reflections are visible in the experimental resolution. The X-ray diffraction pattern confirms the (001) orientation of YBCO film.}
\end{figure*}
\begin{figure*}[h]
\includegraphics[width=14cm]{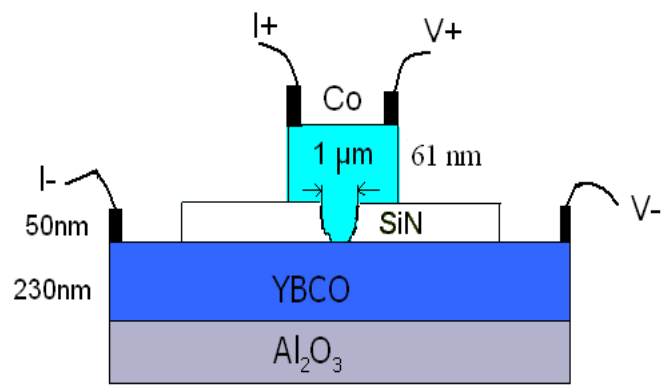}

\caption{\label{fig:epsart}  Schematic diagram of the  Co/YBCO sample with contact geometry. Here, the thickness of Co film near the hole region  is around 61 nm. The hole diameter is 1 $\mu$m on the top  and it is much smaller at the bottom. }
\end{figure*}

\begin{figure*}[h]
\includegraphics[width=10cm]{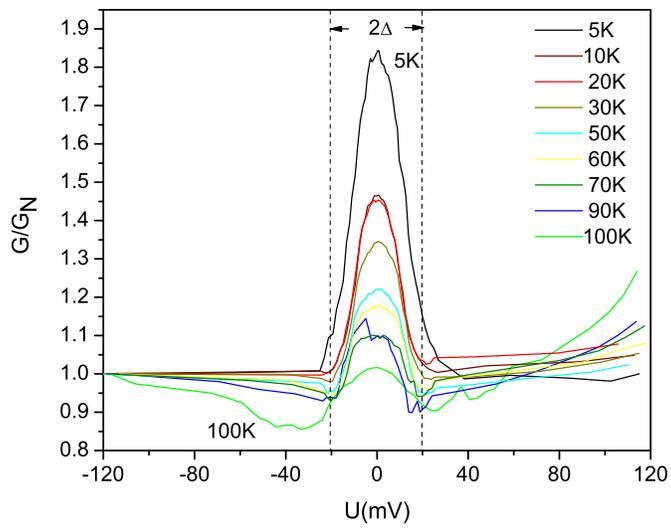}

\caption{\label{fig:epsart} ( On line colour) Normalized conductivity data of Co/YBCO  from 5~K to 100~K. Two vertical dashed lines indicate the 2$\Delta$  superconducting gap  region in  YBCO.}
\end{figure*}

\begin{figure*}[h]
\includegraphics[width=10cm]{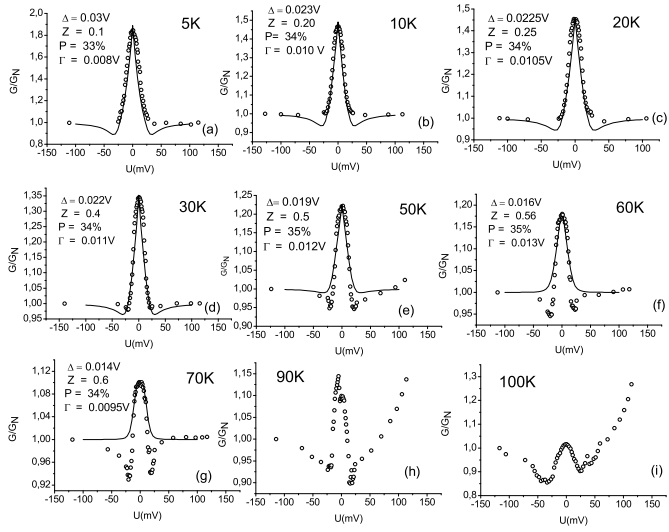}
\caption{\label{fig:epsart} Experimental and modelled normalized
conductivity data   for Co/YBCO with contact area of 1 $\mu$m diameter  at
 (a)10~K, (b)20~K, and (c)100~K. The values of fitted parameters are shown  in the legends
together with error.  Two vertical dashed lines indicate  the 2$\Delta$ superconducting gap region. }

\end{figure*}

\begin{figure*}[h]
\includegraphics[width=10cm]{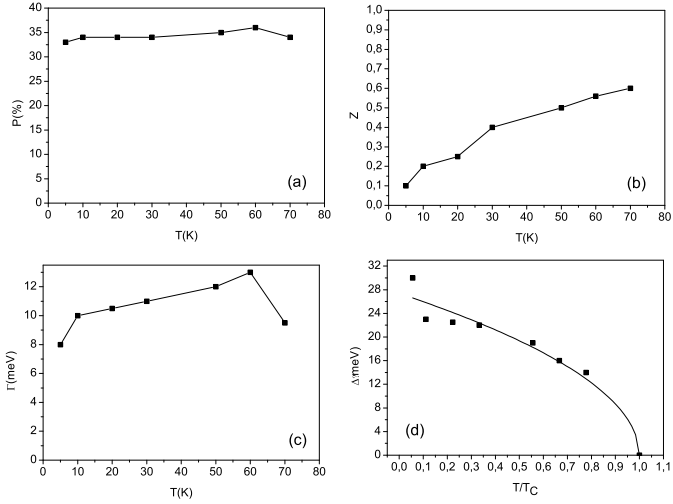}
\caption{\label{fig:epsart} Temperature dependent fitting parameters  for the Co/YBCO sample with
 (a)  the spin polarization, $\it{P}$  (b)  the interface quality, $\it{Z}$ and  the broadening parameter,$\it{\Gamma}$   (c)  the superconducting gaps,$\Delta_{1}$ and $\Delta_{2}$. In (a) and (b) the points are fitting parameters and solid lines are guide to the eye. In (b) the closed and open symbols represent $\it{Z}$ and $\it{\Gamma}$ respectively. In (c)   the open and closed symbols label  $\Delta_{1}$ and $\Delta_{2}$ respectively. Solid lines are fits according to BCS theory. $\Delta_{0}$ = 20 meV is the superconducting gap for YBCO at T=0~K. }

\end{figure*}

\begin{figure*}[h]
\includegraphics[width=10cm]{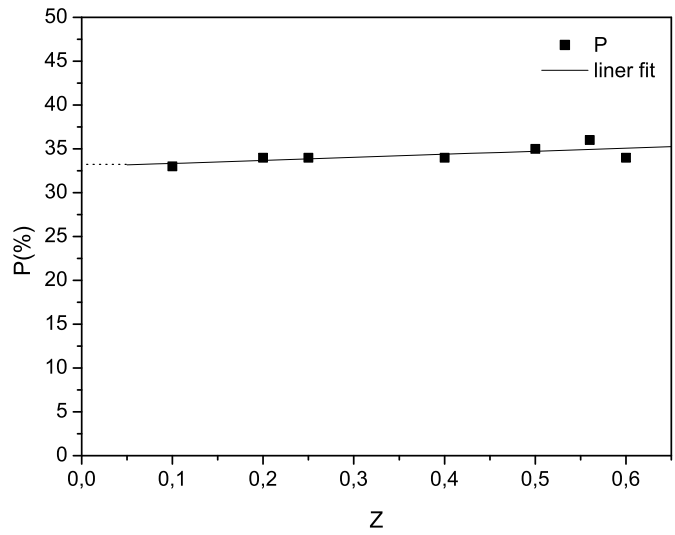}
\caption{ Spinpolarization vs the interface quality, $\it{Z}$  for Co/YBCO }

\end{figure*}

\end{document}